\input{aipcheck}

\documentclass[
    ,final            
  ]
  {aipproc}

\layoutstyle{6x9}

\newcommand{\ket}[1]{| {#1} \rangle}

\usepackage{amsmath}

\begin{document}

\title{\bf Stochastic approach to correlations beyond the mean field
with the Skyrme interaction}


\author{Y.~Fukuoka}{
  address={Institute of Physics, University of Tsukuba, Tsukuba, 305-8571,
           Japan}
}
\author{T.~Nakatsukasa}{
  address={RIKEN Nishina Center, Wako-shi 351-0198, Japan}
  ,altaddress={Center for Computational Sciences, University of Tsukuba,
           Tsukuba 305-8571, Japan}
}
\author{Y.~Funaki}{
  address={RIKEN Nishina Center, Wako-shi 351-0198, Japan}
}
\author{K. Yabana}{
  address={Center for Computational Sciences, University of Tsukuba,
           Tsukuba 305-8571, Japan}
  ,altaddress={RIKEN Nishina Center, Wako-shi 351-0198, Japan}
}

\begin{abstract}
Large-scale calculation based on the multi-configuration Skyrme density
functional theory is performed
for the light $N=Z$ even-even nucleus, $^{12}$C.
Stochastic procedures and the imaginary-time evolution are utilized to
prepare many Slater determinants.
Each state is projected on eigenstates of parity and angular momentum.
Then, performing the configuration mixing calculation with the
Skyrme Hamiltonian, we obtain low-lying energy-eigenstates and their
explicit wave functions.
The generated wave functions are 
completely free from any assumption and symmetry restriction.
Excitation spectra and transition
probabilities are well reproduced,
not only for the ground-state band,
but for negative-parity excited states and
the Hoyle state.
\end{abstract}

\maketitle


\section{Introduction}
\label{sec:intro}

Density functional theory is currently regarded as the only tractable
theory that can be applied across the entire chart of nuclei.
Since nuclei are self-bound systems, they produce their own effective
confining potential (``mean field'').
The mean field often leads to a symmetry-breaking state
which represents certain correlation.
However, for finite nuclei, quantitative description often requires the
symmetry restoration and correlations beyond the mean field.

For light nuclei,
it is important to take into account the cluster correlations,
since a variety of cluster states are known to emerge in excited states
at relatively low energy.
Several theoretical approaches were developed for this \cite{IHS80},
assuming cluster configurations, such as the
resonating group method (RGM), the generator coordinate method (GCM),
and the orthogonal condition model (OCM).
In these approaches, the existence of clusters is assumed from the
beginning.

Therefore, it is desirable to develop a method which is able to
describe both full mean-field dynamics and
correlations beyond mean field,
including the cluster correlation, shape fluctuation, and shape mixing.
For this purpose, in Ref.  \cite{SONY06},
we proposed a new approach and applied it to
the simple BKN energy functional of Ref. \cite{BKN76}.
The method is somewhat similar to the GCM, however,
has an advantage over it.
Namely, since we do not assume a generator coordinate (collective
space) a priori,
the calculation, in principle, may take account of all the possible
correlations beyond the mean field.
Its apparent drawback is necessity of heavy computational resources.

\section{Brief description of method}
In this section, we briefly recapitulate our method.
Readers should be referred to Ref.\cite{SONY06} for details.

We start with a Skyrme Hamiltonian, in the present work,
with the SLy4 parameter set.
Since the Skyrme interaction is an effective interaction
with zero-range forces, it does not have a
real (physical) ground state in the infinite Hilbert space.
Thus, we must define the space consistent with the interaction.

For this purpose, we adopt stochastic generation and imaginary-time
evolution of Slater determinants \cite{SONY06}.
Initial Slater determinants are generated by using random numbers.
However, these initial states contain high momentum components which are
not consistent with the Skyrme interaction.
This motivates us to use the imaginary-time method which can effectively
remove these high-momentum components in the wave function.
Thus, we simply discard Slater determinants produced by the imaginary-time
propagation until the state is sufficiently cooled down.

The imaginary-time propagation produces a kind of ``collective paths''
which are associated with low-energy modes of excitation, 
eventually leads to the Hartree-Fock ground state.
During this imaginary-time propagation, we regularly store the generated
Slater determinants and check their linear independence according to
a certain criterion \cite{SONY06}.
The Slater determinants, $\ket{\Phi_i}$, are constructed in the
mesh representation of the three-dimensional (3D) Cartesian coordinate space.
Then, we perform the parity and the 3D angular-momentum projection,
$\ket{\Phi_{i}^{\pi IK}}_M=P_{MK}^I \ket{\Phi_i}$.
Finally we obtain the energy eigenstate,
$\ket{\Psi^{\pi I}}=\sum_{iK} c_{iK}^{\pi I} \ket{\Phi_i^{\pi IK}}$,
by solving the generalized eigenvalue equation in
each parity and angular momentum sector:
\begin{equation}
\sum_{i'K'}\left(  {\cal H}_{iK,i'K'}^{\pi I}
- E {\cal N}_{iK,i'K'}^{\pi I} \right) c_{i'K'}^{\pi I} =0.
\end{equation}
Here, ${\cal H}_{iK,i'K'}^{\pi I}$
and ${\cal N}_{iK,i'K'}^{\pi I}$ are the Hamiltonian and norm kernels
\cite{RS80}.
In order to avoid the overcompleteness problem
in the configuration mixing calculation,
if the smallest eigenvalue of the norm matrix ${\cal N}_{iK,i'K'}^{\pi I}$
is smaller than $10^{-3}$,
we reduce the number of basis.

The most time-consuming part of the calculation is the angular-momentum
projection \cite{OYN04}.
Since the wave functions are completely free from symmetry restriction,
we must perform the full 3D rotation with respect to three Euler angles.
This becomes more and more difficult as increasing the angular momentum.
Because of this numerical difficulty, we only calculate low-spin states
with $J<6$.
The most delicate part of the calculation is treatment
of the overcomplete basis.
This is discussed in detail in Ref.~\cite{SONY06,Fukuoka12}.

\section{Results}
We apply the method to calculation of excitation spectra in $^{12}$C.
The space is the sphere of the radius 8 fm which is discretized in mesh
of $\Delta x = \Delta y = \Delta z = 0.8$ fm.
In this paper, we show results with the SLy4 parameter set, however
we confirm almost identical excitation spectra with the SkM*, SGII,
and SIII parameters
(although the absolute binding energy of the ground state is different).

\begin{figure}[t]
\centerline{
\includegraphics[width=10cm]{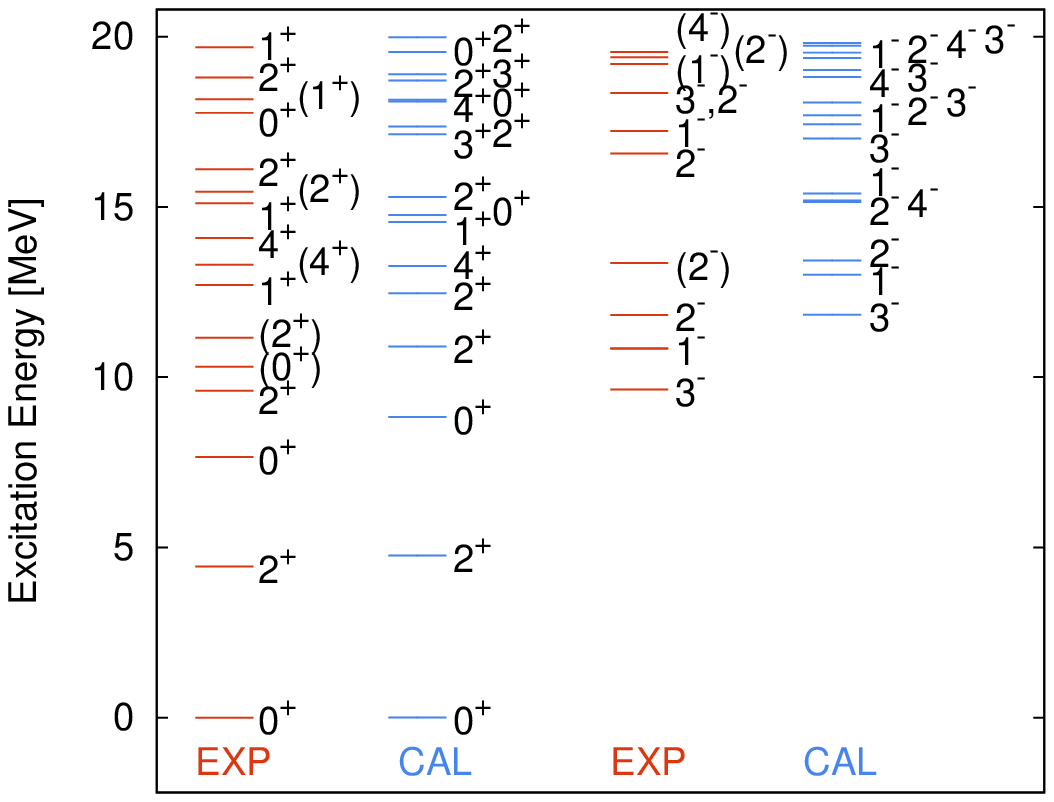}
}
\caption{Excitation spectra in $^{12}$C.
The positive-parity and negative-parity states are shown
in the left and right parts, respectively.
The experimental data are taken from Refs. \cite{Ajz90,Freer09,Freer11}.
}
\label{fig:12C}
\end{figure}

In Fig.~\ref{fig:12C}, we show the result.
In the positive-parity sector, the calculation successfully produces 
the ground-state band ($J^\pi=0^+$, $2^+$, and $4^+$ states).
This rotational character comes from correlation beyond the mean field,
because the Hartree-Fock ground state calculated with SLy4 parameter set 
has a spherical shape.
$B(E2)$ transition strengths among these three states are calculated as
$B(E2;2_1^+\rightarrow 0_1^+)=8.6$ e$^2$fm$^4$ and
$B(E2;4_1^+\rightarrow 2_1^+)=13.4$ e$^2$fm$^4$.
Experiments \cite{Ajz90} indicate
$B(E2;2_1^+\rightarrow 0_1^+)=7.6\pm0.4$ e$^2$fm$^2$
which is well reproduced in the present calculation.

The calculation also reproduces the second $0^+$ state known
as the Hoyle state, that plays a crucial role in the stellar Helium
burning process.
This state has recently been attracting attention as a
condensed state of three-alpha particles \cite{Funaki03}.
In our calculation,
this state is described as a superposition of many Slater determinants.
However, the largest component actually exhibits an spatially extended
structure of three alpha particles.
The transition between the Hoyle state and the first $2^+$ state was
measured as $B(E2; 0_2^+ \rightarrow 2_1^+)=13\pm 2$ e$^2$fm$^4$.
The calculated value is 13.6 e$^2$fm$^4$ in good agreement with
the experiment.
The calculation also well reproduces the $E0$ transition matrix element
between the ground state and Hoyle state;
$M(E0)_{\rm cal}=4.5$ fm$^2$ for the experimental value,
$M(E0)_{\rm exp}=5.4\pm 0.2$ fm$^2$.

The calculated third $0^+$ state at $E_x\approx 15$ MeV has
a very elongated structure,
which may be regarded as a linear chain of the three alpha's.
This state has a large $B(E2)$ transition strength to $2_4^+$ state.

Negative-parity states are also reasonably well reproduced in the calculation,
though the excitation energy is too high by a few MeV.
We obtain the correct ordering among different $J$ multiplets.
It turns out that major components of the low-lying negative parity
states ($3_1^-$, $1_1^-$, and $2_1^-$ states) are
the three-alpha structure, but their spatial extension is smaller
than that of the $0_2^+$ state.


\section{Conclusion}
We have performed the 3D real-space calculations
of the multi-configuration Skyrme density
functional theory,
using the stochastic generation and the imaginary-time propagation of
Slater determinants.
We solve the generalized eigenvalue problem in a generated model space
after the parity and angular-momentum projection.
Low-lying spectra in $^{12}$C are well reproduced.
Properties of the ground-state band,
which consists of $0_1^+$, $2_1^+$, and $4_1^+$ states,
are almost perfectly reproduced.
The Hoyle state ($0_2^+$ state) also shows a nice agreement with experiment.
It is known that the Hoyle state cannot be described in a small model space,
because of its spatially extended character.
The satisfactory description of the Hoyle state indicates that the 
present approach in the 3D coordinate-space representation is able to
take into account correlations beyond the mean field
in a very large model space.

\begin{theacknowledgments}
The work is supported by KAKENHI (Nos. 21340073 and 20105003) and
by SPIRE, MEXT, Japan.
The numerical calculations were performed on
T2K supercomputers in University of Tsukuba.
\end{theacknowledgments}



\bibliographystyle{aipproc}   

\bibliography{myself,nuclear_physics,current,chemical_physics}

\IfFileExists{\jobname.bbl}{}
 {\typeout{}
  \typeout{******************************************}
  \typeout{** Please run "bibtex \jobname" to optain}
  \typeout{** the bibliography and then re-run LaTeX}
  \typeout{** twice to fix the references!}
  \typeout{******************************************}
  \typeout{}
 }

\end{document}